\documentstyle{article}
\newcommand{\bk}{{\bf k}}
\newcommand{\br}{{\bf r}}
\newcommand{\beq}{\begin{equation}}
\newcommand{\eeq}{\end{equation}}
\newcommand{\beqa}{\begin{eqnarray}}
\newcommand{\eeqa}{\end{eqnarray}}

\newcommand{\om}{\omega }
\newcommand{\Om}{\Omega }
\def \l {\left}
\def \r {\right}

\begin{document}
\title{   Conductivity in Two-Dimensional Disordered
Model with Anisotropic Long-Range Hopping.}

\author{E. A. Dorofeev, S. I. Matveenko }
\date{}
\maketitle
{ \it Landau Institute for Theoretical Physics, Moscow,
Kosygina str., 2, Russia, 117940}

\vskip .1in
\begin{abstract}
We consider two-dimensional system of particles
localized on randomly distributed sites of squared lattice
with anisotropic transfer matrix elements between localized sites.
By summing of "diffusion ladder" and "cooperon ladder" type
vertices we calculated the conductivity for various sites and
particles densities.
The model is relevant to the problem of strong nonmagnetic impurities
in superconductors with $d_{x^2 -y^2}$ symmetry of the  order
parameter \cite{bal}.


\end{abstract}

\section{ Introduction}

 The following tight-binding Hamiltonian is considered:
\beq
H= \sum_{i,j} t({\bf r}_j - {\bf r}_i )\psi^{+}({\bf r}_i)
\psi^{+}({\bf r}_j)
\rho ({\bf r}_i)\rho ({\bf r}_j),
\label{ham}
\eeq
where $\psi^{+}({\bf r}_i), \psi^{+}({\bf r}_j)$ are creating and
annihilation operators, $\rho ({\bf r}_i)$ is the filling number,
equal to 1 at
the localized sites, and to 0  otherwise.
The transfer matrix element has a cross-shape
configuration
\beq
 t({\bf r}) = (\delta_{x,0} + \delta_{y,0}) f(r),
\label{t}
\eeq
 with
$$f(r) = J\ \l ( \frac{a}{r} \r )^{\gamma } \exp{(-\kappa r)},$$
and $a$ is a lattice constant.
We examine the case of a random distribution of impurities on a sites of
two-dimensional squared lattice. An impurity potential generates a
localized state with strongly anisotropic wave function. As a result
 the conductivity is carried out due to hoppings of particles between
local states on the same vertical or horizontal lines.
A similar picture can be realized in 2D  $d_{x^2 - y^2}$-wave
 superconductors, where local bound quasiparticle states may
arise in the presence of unitary impurities \cite{bal}.

The plan of this article is as follows. In Chapter 2 we consider
the case of low impurity density. In
Chapter 3  we calculate the conductivity in the case of high impurity
density. In the Conclusion we
discuss our results.

\vskip.1in
\section{Low density}

We consider now the limit of low impurities concentration ($c\ll 1 $).
In the case of an external electromagnetic field we should
substitute in (\ref{ham})
$$ t(\br_i -\br_j ) \to t(\br_i -\br_j )
\exp({ie}\int_{\br_i}^{\br_j}{\bf A}(\br, t)d\br ).
$$
The electric current is defined as usually from
the Hamiltonian (\ref{ham}) by
varying over a gauge invariant vector-potential ${\bf A}$
\beq
j_{\alpha}(t) = -i e \sum_{i,j} (\br_i -\br_j )_{\alpha} t_{ij}
\Psi_i^{+}(t) \Psi_j(t) \rho_i
\rho_j \exp({ie}{\bf A }(t)({\br_i} -{\br_j})).
\label{j1}
\eeq

Where $ t_{ij} = t( \br_i - \br_j) $.
Since we will calculate $j(\omega )$ we consider that the potential
${\bf A}$ depends on time $t$ only.
Using the equation for the Green function $G(t_1,\br_1,t_2, \br_2 )$
\beq
\frac{\partial}{\partial t_1 } G(t_1,\br_1,t_2, \br_2) =
-i<T \frac{\partial \Psi (1)}{\partial t_1 } \Psi^+ (2)> -i\delta
(\br_1 -\br_2)
\delta (t_1 -t_2 )
\eeq
we obtain after Fourier transformation in the linear over
${\bf A }$ approximation
\beqa
j_{\alpha}(\om )= \frac{e^2}{c}\sum_{i,j}t_{i,j}(\br_i -\br_j )_{\alpha}
(\br_i -\br_j)_{\beta} A_{\beta }(\om)
\int \frac{d\Om }{2\pi }G(\Om ,\br_j, \br_i )e^{i\Om \alpha } -
\nonumber \\
\frac{e^2}{c}\sum_{i,j,k,l}t_{i,j}t_{k,l}(\br_i -\br_j )_{\alpha}
(\br_k -\br_l )_{\beta} A_{\beta }(\om)
 \int \frac{d\Om }{2\pi }G(\om +\Om ,\br_j, \br_k )G(\Om ,\br_l, \br_i )
e^{i\Om \alpha }.
\label{j2}
\eeqa
The summation in eqn.(\ref{j2}) is taken over impurity sites, and
$\alpha \to +0$.
In order to evaluate (\ref{j2}) in the lowest order with respect
to the concentration we examine the case of two random arranged sites.
The Green function is found easy
\beqa
G(\om , \br_i, \br_i )= \frac{\om + \mu }{(\om +\mu )^2 - t^2_{i,j}},
\nonumber \\
G(\om , \br_i, \br_j )= \frac{t_{1,2}}{(\om +\mu )^2 - t^2_{i,j}},
\label{g}
\eeqa
where $\mu $ is the chemical potential.
Substituting (\ref{g}) to (\ref{j2}) we obtain
in the case when sites are arranged on the same horizontal chain
\beq
j_x(\om ) = \frac{e^2}{2c}(x_1 -x_2)^2 t_{1,2}
\frac{\om^2 }{(\om + i\alpha )^2 -4t_{1,2}^2 }A_x(\om )
= Q(\om )A_x(\om ) .
\label{j3}
\eeq
The similar equation can be derived for the case of nonzero
temperatures. The result differs only by Fermi filling factor.
After averaging over impurities sites we obtain for the conductivity
$\sigma (\om ) = iQ(\om ) / \om $
\beqa
\sigma (\om ) = \frac{\pi e^2}{4} c^2 L \int x^2 t(x)[n_F (\om -\mu )-
n_F (-\om -\mu )]\delta (\om - 2 t(x))dx \nonumber \\
= \frac{\pi e^2}{8} c^2 L \frac{x_0^2 t(x_0 ) [n_F (\om -\mu )-
n_F (-\om -\mu )]}{\vert t^{\prime }(x_0 )\vert },
\label{s}
\eeqa
where $n_F $ is the Fermi distribution function, $2t(x_0 ) =\om $.

Substituting $ t(x) $ from (\ref{t}) we get
\beq
\frac{\om }{2J} = \frac{\exp(-\kappa x_0 )}{x_0^{\gamma }}.
\label{x}
\eeq
\beq
\sigma (\om ) = \frac{\pi e^2}{8} c^2 L\frac{x_0^3 [n_F (\om -\mu )-
n_F (-\om -\mu )]}{\gamma +\kappa x_0 }.
\label{s2}
\eeq
In the limit of low frequency we have following asymptotic behaviors

\noindent 1. $\kappa = 0$, $x_0 =(2J/\om )^{1/\gamma }$:
\beq
 \om \gg T, \sigma (\om ) \propto \om ^{-3/\gamma },
\label{}
\eeq
\beq
\om \ll T, \sigma (\om ) \propto \om ^{-3/\gamma +1},
\label{}
\eeq

\noindent 2. $\kappa \neq 0$, $\kappa x_0 \sim \log (2J/\om )$:
\beq
\om \gg T, \sigma (\om ) \propto \log^2 \frac{2J}{\om },
\label{}
\eeq
\beq
\om \ll T, \sigma (\om ) \propto \om \log^2 \frac{2J}{\om}.
\label{}
\eeq

\section{High density.}
\subsection{ Green function.}

In the case of high density of impurities we will assume that
distribution function of impurities may be approximate by Gausses
distribution  with dispersion $g$. That is
\beqa
\rho({\bf r}_i) = c +  \delta \rho(\br_i)  \nonumber \\
<\delta \rho ({\bf r}_i) \delta \rho ({\bf r}_j)>_{\rho}= g^2 \delta _{ij}.
\label{G}
\eeqa
We will assume that concentration $c  \leq 1$.

The  one-particle Green function for the arbitrary impurity
distribution is defined  in terms of functional integral as usually
\beqa
G(E, {\bf r},{\bf r}^{\prime}) = (E - t_{ij}\rho ({{\bf r}_i})
\rho ({\bf r}_j) )^{-1}_{{\bf r},{{\bf r}^{\prime}}} = \nonumber \\
i\frac{\int D\bar {\psi} D{\psi}\psi ({\bf r}){\bar \psi}({\bf r}^{\prime})
\exp (iS)}{\int D\bar{\psi} D{\psi} \exp (iS) },
\label{gr}
\eeqa
where
\beq
S = S_0+S_1,
\label{}
\eeq

\beq
iS_0 = i \sum_{\br} \bar{\psi(\br)} E \psi(\br),
\label{}
\eeq

\beq
iS_1 = -i \sum_{\br_1 \br_2} \bar{\psi}(\br_1) t(\br_1-\br_2)
\rho(\br_1) \rho(\br_2) \psi(\br_2).
\label{}
\eeq
Introducing an additional integration over new fields $\chi$ , $\bar{\chi}$
in order to eliminate the second order terms $\rho \rho$, we get

\beqa
e^{iS_1}&=& \int D \bar{\chi} D \chi \exp \{ -i \sum \bar{\chi(\br_1)}
t^{-1}(\br_1-\br_2) \chi(\br_2)
\nonumber \\
&& + c \sum_{\br}( \bar{\chi}(\br) \psi(\br) + \bar{\psi}(\br)
\chi(\br)) \nonumber \\
&& +\sum_{\br} \delta \rho(\br) ( \bar{\chi}(\br) \psi(\br) +
\bar{\psi}(\br) \chi(\br)) \}/Z
\label{}
\eeqa
where
\beq
Z = \int D \bar{\chi} D \chi \exp \{-i \sum_{\br_1 \br_2} \bar{\chi}
(\br_1) t^{-1}(\br_1-\br_2) \chi(\br_2)\}
\eeq
\beq
t^{-1}(\br) = \frac{1}{V} \sum_{\bk} \varepsilon^{-1}(\bk)
e^{i \bk \br}
\label{}
\eeq

\beq
\varepsilon(\bk) = \sum_{\br}  t(\br) e^{i \bk
\br }= - J \ln [ \l (\kappa^2+4 \sin^2 \l ( \frac{k_x a}{2} \r )
\r ) \l ( \kappa^2 + 4 \sin^2 \l ( \frac{k_y a}{2} \r ) \r) ]
\label{}
\eeq

The Green function in terms of new two component field  $\varphi $

\beqa
\varphi_1(\br)& =& \psi(\br)  \nonumber \\
\varphi_2(\br)& = & \chi(\br)  \nonumber \\
\bar{\varphi}_1(\br)& =& \bar{\psi}(\br)  \nonumber \\
\bar{\varphi}_2(\br)& =& \bar{\chi}(\br)
\label{}
\eeqa
reads

\beq
{\hat G}(\br_1, \br_2) = -i < \hat{\varphi}(\br_1) \otimes
\hat{\bar{\varphi}}(\br_2) >
\label{}
 \eeq
where  $\hat{\bar{\varphi}}(\br )  = (\bar{{\varphi}}_1(\br),
\bar{\varphi}_2 (\br ))$
and
angle brackets are defined as
\beq
< \ldots > = \frac {\int D \bar{\psi} D \psi D \bar{\chi} D \chi (\ldots)
e^{iS_{eff}}} {\int D \bar{\psi} D \psi D \bar{\chi} D \chi e^{iS_{eff}}}
\label{}
 \eeq
with
\beq
iS_{eff} =\sum_{\bf r }\hat{\bar{\varphi}}(\br )\left( \begin{array}{cc}
                             iE & c+  \delta \rho(\br) \\
 c+  \delta \rho(\br)  & -i t^{-1}(\br_1-\br_2) \\
\end{array} \right) \hat{\varphi}(\br ).
\label{iS}
\eeq

The  equation for the Green function after averaging over impurities
in the Born approximation reads
\beq
{\hat G}(\bk) = {\hat G}^{0}(\bk) + {\hat G}^{0}(\bk) {\hat \Sigma}(\bk)
{\hat G}(\bk)
\label{d}
\eeq
with the bare Green function
\beq
[{\hat G}^0(\bk)]^{-1} = \left [
\begin{array}{ll}
E & -ic \\
-ic & -\varepsilon^{-1}(\bk)
\end{array}
\right ]
\label{}
 \eeq
and the self-energy ${\hat \Sigma} $  obtained by summing of diagrams without
intersections
\beq
{\hat \Sigma}(\bk) = g^2 a^2 \int \frac{ d \bk_1}{(2\pi)^2 } \sigma^x {%
\hat G}(\bk_1) \sigma^x.
\label{si}
\eeq

The solution of Eqns. (\ref{d}), (\ref{si}) is
\beq
\Sigma(\bk) = \left (
\begin{array}{cc}
Q & -iP \\
-iP & R
\end{array}
\right )
\label{} \eeq

\beq
{\hat G}(\bk) = \frac{1}{(1+ \varepsilon(\bk) R)(E-Q) -(c+P)^2
\varepsilon(\bk)} \left [
\begin{array}{cc}
1+ \varepsilon(\bk) R & -i(c+P) \varepsilon(\bk) \\
-i(c+P) \varepsilon(\bk) & -\varepsilon(\bk)(E-Q)
\end{array}
\right ]
\label{} \eeq
where
\beqa
Q & = & g^2 a^2 \int \frac{ d \bk }{(2\pi)^2 } \frac{(E-Q) \varepsilon(%
\bk)} {(1 + \varepsilon(\bk) R)(E-Q) -(c+P)^2 \varepsilon(%
\bk) } \nonumber\\
R & = & - g^2 a^2 \int \frac{ d \bk }{(2\pi)^2 } \frac{1 - \varepsilon(%
\bk) R} {(1 + \varepsilon(\bk) R)(E-Q) -(c+P)^2 \varepsilon(%
\bk) } \nonumber\\
P & = & g^2 a^2 \int \frac{ d \bk }{(2\pi)^2 } \frac{(c + P)\varepsilon(\bk)}
 {(1 +\varepsilon(\bk) R)(E-Q) -(c+P)^2 \varepsilon(\bk) }. \nonumber
\label{} \eeqa

In the limit of low dispersion
$g^2 \ll 1$ we obtain

\beqa
Q^{R,A} & = & \pm i \frac{\gamma}{2} \nonumber \\
R^{R,A} & = & \mp i  \frac{c^2}{E^2} \frac{\gamma}{2} \\
P^{R,A} & = & \pm i   \frac{c}{E} \frac{\gamma}{2} \nonumber
\eeqa

\beq
\gamma = 2\pi g^2 a^2 \frac{E^2}{c^4} \nu \l ( \frac{E}{c^2} \r )
\label{} \eeq
Where $\nu_0(\varepsilon)$ is density of states of the pure model
(c=1, g = 0):
\beq
\nu (\varepsilon) = \int \frac{d \bk}{(2\pi)^2} \delta(\varepsilon -
\varepsilon(\bk)).
\label{} \eeq
Taking into account that
$Q \ll E$, $RE \ll 1$, $P \ll c$,
we find for the Green function in the limit $g^2 \ll 1 $

\beq
{\hat G}^{R,A}(\bk) = \frac{1}{E - c^2 \varepsilon(\bk) \pm i
\gamma/2} \left (
\begin{array}{cc}
1 & -ic \varepsilon(\bk) \\
-ic \varepsilon(\bk) & -\varepsilon(\bk) E
\end{array}
\right )
\label{bgr}
 \eeq

\subsection{Drude formula}

The conductivity in our case is defined as in Ch.I in terms of
four-particle correlation function

\beq
\sigma _{E}(\omega) = \frac{e^2}{2\pi } \int \frac{d \bk_1}{(2\pi)^2}
\frac { d \bk_2 }{(2\pi)^2} v_{ \alpha }( \bk_1 ) v_{ \alpha } (
\bk_2) K_{E \omega}( \bk_1, \bk_2; \bk_2,
\bk_1)
\label{cond}
\eeq
where $E$ is taken at the Fermi level,
\beqa
K_{E \omega}( \bk_1, \bk_2; \bk_2, \bk_1) =
\nonumber \\
\frac{1}{V} \sum_{{\bf x,y,z,t}} e^{i \bk_1 ({\bf x}-{\bf y})}
e^{i \bk_2 ({\bf z}-{\bf t})}
\langle \rho_{{\bf x}} \rho_{{\bf y}} \rho_{{\bf z}} \rho_{{\bf t}}
G^R_{11}({\bf y},{\bf z},E+\omega/2)
G^A_{11}({\bf t},{\bf x},E-\omega/2) \rangle_{\rho}
\label{}
\eeqa
and
\beq
v_{ \alpha }( \bk ) = \frac {\partial \varepsilon ( \bk)}{
\partial k_{\alpha}}
\label{}
\eeq

Substituting the solution (\ref{bgr}) to (\ref{cond})
we find in the lowest approximation

\beq
\sigma _{E}(\omega) = \frac{e^2}{2\pi } \int \frac{d \bk}{(2\pi)^2}
 v^2_{ \alpha }(\bk)G^R_{11}(\bk,E+\omega/2) G^A_{11}(\bk,E-\omega/2)=
 \frac{e^2 c^6}{2\pi}\l( \frac{J}{E} \r)^2
\frac{A(E)}{B(E)}
\label{cond}
\eeq
where
\beq
A(E) = \int_{ \varepsilon(\bk) = E/c^2} \frac{dl_{\bk}}{v(\bk)}
v_{x}^{2}(\bk), \qquad
B(E) = \int_{ \varepsilon(\bk) = E/c^2} \frac{dl_{\bk}}{v(\bk)},
\eeq
 $v(\bk) = \sqrt{v_{x}^{2}(\bk)+v_{y}^{2}(\bk)}$,and
$dl_{\bk}$ - element of the length of the Fermi surface.

The conductivity can be expressed in terms of  a particle density defined as
\beq
n_{0}(E) = a^2 \int_{-\pi/a}^{\pi/a} dk_{x}  \int_{-\pi/a}^{\pi/a} dk_{y}
\theta( E/c^2 - \varepsilon(\bk)).
\eeq
We obtain in limits of low and high densities following results:

\beq
\sigma = \left\{ \begin{array}{c}
 \frac{e^2 c^6}{g^2} \frac{1}{32 \pi \log^{2}(2)}n_{0}, \qquad
 {\rm for }\quad n_0  \ll 1, \\
 \frac{e^2 c^6}{g^2} \frac{1}{4 \kappa^4
\log^{2}(\kappa)}(1-n_0 ), \qquad  {\rm for } \quad  (1- n_0 ) \ll 1. \\
\end{array} \right.
\eeq
the asymptotic behavior in intermediate region  $ 0 < n_0 < 1$ is
\beq
 \sigma = 2 \pi^4 \frac{e^2 c^6}{g^2} \frac{1}{(1-n_0 )^2
\log \l ( \frac{1}{1 - n_0 } \r)}
\eeq
with maximum value
\beq
\sigma_{max} \sim \frac{1}{ \kappa^{8/3} \log(\kappa)}
\eeq
 reached at $  1- n_0  \sim \kappa^{4/3} $.

\subsection{The absence  of weak localization.}
To go beyond the quasiclassical approximation we include contributions to
the conductivity from "diffusion-ladder" and "crossed-ladder" or "cooperon"
vertices \cite{gor, abr}. 
 The addition term is

\beq
\delta \sigma _{E}(\omega) =
\frac{e^2}{2\pi } \int \frac{d \bk_1}{%
(2\pi)^2} \frac { d \bk_2 }{(2\pi)^2} v_{ \alpha }( \bk_1 ) v_{
\alpha } ( \bk_2) G^R_{1a}(\bk_1) G^R_{c1}(\bk_2)
G^A_{1b}(\bk_2) G^A_{d1}(\bk_1) K_{ac;bd}( \bk_1,\bk_2)
\label{ad}
\eeq
where for "diffusion" - vertex contribution we get the
equation (see Fig.1a):
\beq
K^{(D)}_{ac;bd} = g^2 \sigma^x_{ac} \sigma^x_{bd} + g^2 \int \frac{d
{\bf l}}{(2\pi)^2}
\sigma^x_{a a_1}G^{R}_{a_1 c_1}({\bf l})
K^{(D)}_{c_1 c ; b b_1 } G^{A}_{b_1 d_1}({\bf l}) \sigma^x_{d_1 d},
\label{D} \eeq
The solution of this equation  does not depend on $\bk_1$ and $\bk_2$.
 Therefore the contribution of a "diffusion" - vertex to the conductivity
is equal to zero because 
\beq
\int \frac{d \bk_1}{(2\pi)^2} v_{ \alpha }( \bk_1 )
 G^R_{1a}(\bk_1)  G^A_{d1}(\bk_1) = 0.
\eeq
\unitlength 1.00mm
\linethickness{0.4pt}
\begin{picture}(145.00,74.00)
\put(9.67,66.34){\vector(1,0){0.2}}
\put(7.34,66.34){\line(1,0){2.33}}
\put(44.00,66.34){\vector(1,0){0.2}}
\put(9.67,66.34){\line(1,0){34.33}}
\put(20.67,66.34){\line(0,-1){19.00}}
\put(34.00,66.34){\line(0,-1){19.00}}
\put(27.00,56.01){\makebox(0,0)[cc]{K}}
\put(49.00,57.01){\makebox(0,0)[cc]{=}}
\put(58.67,66.34){\vector(1,0){0.2}}
\put(57.34,66.34){\line(1,0){1.33}}
\put(71.34,66.34){\vector(1,0){0.2}}
\put(58.67,66.34){\line(1,0){12.67}}
\put(64.67,66.34){\line(0,-1){4.67}}
\put(64.67,61.67){\line(0,-1){4.01}}
\put(64.67,57.65){\line(0,-1){3.36}}
\put(64.67,54.30){\line(0,-1){2.70}}
\put(64.67,51.60){\line(0,-1){2.04}}
\put(64.67,49.56){\line(0,-1){1.38}}
\put(64.67,48.18){\line(0,-1){0.84}}
\put(77.67,57.34){\makebox(0,0)[cc]{+}}
\put(85.34,66.34){\vector(1,0){0.2}}
\put(84.00,66.34){\line(1,0){1.34}}
\put(98.00,66.34){\vector(1,0){0.2}}
\put(85.34,66.34){\line(1,0){12.66}}
\put(129.00,66.34){\vector(1,0){0.2}}
\put(98.00,66.34){\line(1,0){31.00}}
\put(91.67,66.34){\line(0,-1){4.67}}
\put(91.67,61.67){\line(0,-1){4.01}}
\put(91.67,57.65){\line(0,-1){3.36}}
\put(91.67,54.30){\line(0,-1){2.70}}
\put(91.67,51.60){\line(0,-1){2.04}}
\put(91.67,49.56){\line(0,-1){1.38}}
\put(91.67,48.18){\line(0,-1){0.84}}
\put(118.67,66.34){\line(0,-1){19.00}}
\put(104.67,66.34){\line(0,-1){19.00}}
\put(111.67,56.34){\makebox(0,0)[cc]{K}}
\put(9.67,29.34){\vector(1,0){0.2}}
\put(7.34,29.34){\line(1,0){2.33}}
\put(44.00,29.34){\vector(1,0){0.2}}
\put(9.67,29.34){\line(1,0){34.33}}
\put(20.67,29.34){\line(0,-1){19.00}}
\put(34.00,29.34){\line(0,-1){19.00}}
\put(27.00,19.01){\makebox(0,0)[cc]{K}}
\put(49.00,20.01){\makebox(0,0)[cc]{=}}
\put(58.67,29.34){\vector(1,0){0.2}}
\put(57.34,29.34){\line(1,0){1.33}}
\put(71.34,29.34){\vector(1,0){0.2}}
\put(58.67,29.34){\line(1,0){12.67}}
\put(64.67,29.34){\line(0,-1){4.67}}
\put(64.67,24.67){\line(0,-1){4.01}}
\put(64.67,20.65){\line(0,-1){3.36}}
\put(64.67,17.30){\line(0,-1){2.70}}
\put(64.67,14.60){\line(0,-1){2.04}}
\put(64.67,12.56){\line(0,-1){1.38}}
\put(64.67,11.18){\line(0,-1){0.84}}
\put(77.67,20.34){\makebox(0,0)[cc]{+}}
\put(114.00,20.01){\makebox(0,0)[cc]{+}}
\put(119.34,29.67){\vector(1,0){0.2}}
\put(118.67,29.67){\line(1,0){0.67}}
\put(140.34,29.67){\vector(1,0){0.2}}
\put(119.34,29.67){\line(1,0){21.00}}
\multiput(123.67,29.67)(0.12,-0.20){21}{\line(0,-1){0.20}}
\multiput(126.10,25.52)(0.12,-0.20){18}{\line(0,-1){0.20}}
\multiput(128.23,21.88)(0.12,-0.20){16}{\line(0,-1){0.20}}
\multiput(130.08,18.73)(0.12,-0.20){13}{\line(0,-1){0.20}}
\multiput(131.63,16.09)(0.11,-0.20){11}{\line(0,-1){0.20}}
\multiput(132.89,13.94)(0.11,-0.18){9}{\line(0,-1){0.18}}
\multiput(133.85,12.29)(0.11,-0.19){6}{\line(0,-1){0.19}}
\multiput(134.53,11.15)(0.12,-0.20){4}{\line(0,-1){0.20}}
\multiput(124.67,10.34)(0.12,0.22){19}{\line(0,1){0.22}}
\multiput(126.88,14.48)(0.11,0.21){17}{\line(0,1){0.21}}
\multiput(128.83,18.12)(0.11,0.21){15}{\line(0,1){0.21}}
\multiput(130.51,21.27)(0.12,0.22){12}{\line(0,1){0.22}}
\multiput(131.93,23.92)(0.11,0.21){10}{\line(0,1){0.21}}
\multiput(133.07,26.06)(0.11,0.21){8}{\line(0,1){0.21}}
\multiput(133.95,27.71)(0.10,0.19){6}{\line(0,1){0.19}}
\multiput(134.57,28.86)(0.11,0.20){4}{\line(0,1){0.20}}
\put(129.67,29.67){\line(0,-1){4.75}}
\put(129.67,24.91){\line(0,-1){4.08}}
\put(129.67,20.83){\line(0,-1){3.41}}
\put(129.67,17.42){\line(0,-1){2.74}}
\put(129.67,14.67){\line(0,-1){2.07}}
\put(129.67,12.60){\line(0,-1){1.41}}
\put(129.67,11.19){\line(0,-1){0.86}}
\put(145.00,19.67){\makebox(0,0)[cc]{+ ...}}
\put(85.00,29.34){\vector(1,0){1.00}}
\put(86.00,29.34){\vector(1,0){22.33}}
\multiput(91.00,29.34)(0.12,-0.20){20}{\line(0,-1){0.20}}
\multiput(93.40,25.36)(0.12,-0.20){18}{\line(0,-1){0.20}}
\multiput(95.52,21.85)(0.12,-0.19){16}{\line(0,-1){0.19}}
\multiput(97.37,18.79)(0.11,-0.19){14}{\line(0,-1){0.19}}
\multiput(98.94,16.18)(0.12,-0.20){11}{\line(0,-1){0.20}}
\multiput(100.24,14.04)(0.11,-0.19){9}{\line(0,-1){0.19}}
\multiput(101.26,12.35)(0.11,-0.18){7}{\line(0,-1){0.18}}
\multiput(102.00,11.11)(0.12,-0.19){4}{\line(0,-1){0.19}}
\multiput(102.47,10.34)(0.10,-0.16){2}{\line(0,-1){0.16}}
\multiput(92.00,10.01)(0.11,0.21){20}{\line(0,1){0.21}}
\multiput(94.29,14.15)(0.12,0.21){17}{\line(0,1){0.21}}
\multiput(96.30,17.79)(0.12,0.21){15}{\line(0,1){0.21}}
\multiput(98.03,20.94)(0.11,0.20){13}{\line(0,1){0.20}}
\multiput(99.50,23.59)(0.12,0.21){10}{\line(0,1){0.21}}
\multiput(100.68,25.73)(0.11,0.21){8}{\line(0,1){0.21}}
\multiput(101.59,27.38)(0.11,0.19){6}{\line(0,1){0.19}}
\multiput(102.22,28.53)(0.11,0.20){4}{\line(0,1){0.20}}
\put(44.67,47.34){\vector(-1,0){0.67}}
\put(44.00,47.34){\vector(-1,0){35.67}}
\put(72.67,47.00){\vector(-1,0){0.67}}
\put(72.00,47.00){\vector(-1,0){13.33}}
\put(129.67,47.00){\vector(-1,0){0.67}}
\put(129.00,47.00){\vector(-1,0){31.00}}
\put(98.00,47.00){\vector(-1,0){12.67}}
\put(44.67,10.34){\vector(-1,0){1.00}}
\put(43.67,10.34){\vector(-1,0){35.00}}
\put(72.00,10.34){\vector(-1,0){0.67}}
\put(71.33,10.34){\vector(-1,0){12.67}}
\put(109.00,10.00){\vector(-1,0){0.67}}
\put(108.33,10.00){\vector(-1,0){22.33}}
\put(141.33,10.00){\vector(-1,0){0.67}}
\put(140.67,10.00){\vector(-1,0){21.33}}
\put(12.33,73.34){\makebox(0,0)[cc]{a)}}
\put(12.00,35.67){\makebox(0,0)[cc]{b)}}
\put(29.33,21.00){\makebox(0,0)[cc]{(C)}}
\put(114.00,58.67){\makebox(0,0)[cc]{(D)}}
\put(29.33,58.00){\makebox(0,0)[cc]{(D)}}
\end{picture}

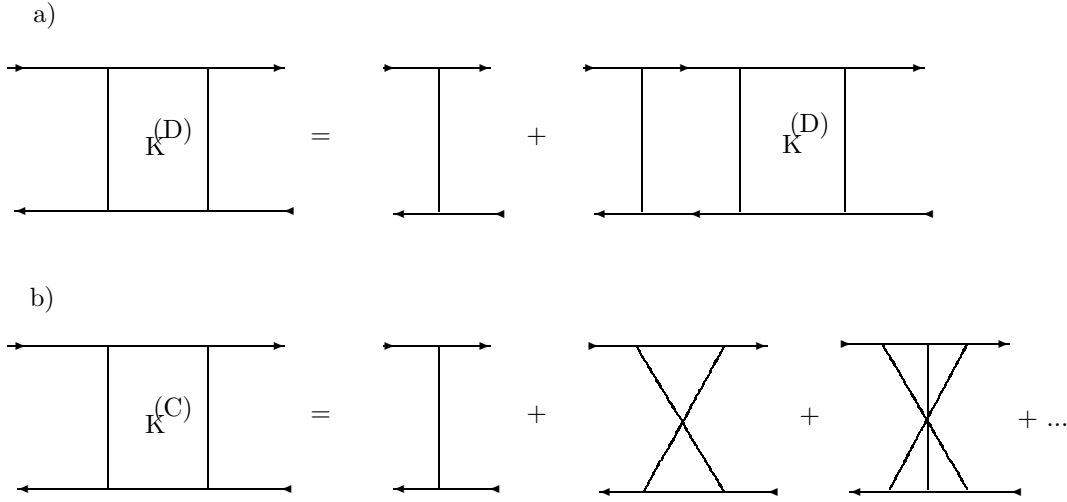
 {\small {\bf Fig.\ 1}. {Diffusion a) and crossed-ladder b)
vertices }.}
\vskip0.5cm

Now we consider the "cooperon"-vertex contribution.
The vertex $K^{(D)}$ obeys  the equation (see Fig. 1b )
\beq
K^{(C)}_{ac;bd}({\bf q}) = g^2 \sigma^x_{ac} \sigma^x_{bd} + g^2 \int
\frac{d
{\bf l}}{(2\pi)^2} \sigma^x_{ae} \sigma^x_{bg} G^{R}_{e a_1}({\bf l})
G^{A}_{g b_1}( {\bf q} - {\bf l}) K^{(C)}_{a_1 c ; b_1 d} ( {\bf q}),
\label{K} \eeq

where
${\bf q} = \bk_1+ \bk_2$.

If the solution of this equation has some singularity behavior as function
of ${\bf q}$ (for example "diffusion pole") we may hope to obtain
the contribution to conductivity. Now we prove that "cooperon"-vertex
has not singularity.

We search the solution of (\ref{K}) in the form
\beq
K^{(C)}_{ac;bd} = g^2 \sum_{ \mu \nu} \sigma^{\mu}_{ac}
\sigma^{\nu}_{bd} A_{\mu
\nu},
\label{}
 \eeq
where $\mu , \nu = 0,x,y,z $, and $A$ satisfies the equation
\beq
A_{ \mu \nu } = \delta_{\mu x} \delta_{\nu x} + \sum_{ \alpha \beta }
\Lambda_{ \mu \nu}^{ \alpha \beta} A_{\alpha \beta}
\label{}
\eeq
with

\beq
\Lambda_{ \mu \nu}^{ \alpha \beta} = \frac{g^2}{4} \int \frac{d {\bf l}}{%
(2\pi)^2} \left [ Sp ( \sigma^x G^R \sigma^{\mu} \sigma^{ \alpha }) \right ]
\left [ Sp ( \sigma^x G^R \sigma^{\nu} \sigma^{ \beta }) \right ]
\label{}
\eeq
The matrix  $\Lambda $ can be rewritten as
\beq
\Lambda_{ \mu \nu}^{ \alpha \beta} = - \Delta^{\alpha}_{\mu}
\Delta^{\beta}_{\nu}
\label{} \eeq
with
\beq
\Delta^{\alpha}_{\mu} = - \frac{1}{2} \frac{c}{E} \left [ Sp ( \sigma^x g
\sigma^{\mu} \sigma^{ \alpha }) \right ]
\label{} \eeq
and

\beq
g = \left [
\begin{array}{cc}
i & E/c \\
E/c & -i E^2/c^2
\end{array}
\right ]
\label{} \eeq
So the components of matrix $\Lambda $  are
\beq
\hat {\Delta} = \left(
\begin{array}{cccc}
1 & i \sinh(\theta) & \cosh(\theta) & 0 \\
i \sinh(\theta) & 1 & 0 & -i \cosh(\theta) \\
\cosh (\theta) & 0 & 1 & -\sinh(\theta) \\
0 & i\cosh(\theta) & \sinh(\theta) & 1
\end{array}
\right)
\label{} \eeq
where
\beq
c/E = e^{\theta}
\label{} \eeq
The solution of (\ref{})  is found in terms of matrices $U$, $B$:

\beq
A_{\mu \nu} = U_{\mu \mu_1} U_{\nu \nu_1} B_{ \mu_1 \nu_1},
\label{} \eeq
where
\beq
(U^{-1})_{\mu \mu_1} \Delta_{\mu_1 \nu_1} U_{ \nu_1 \nu} = z_{\mu}
\delta_{\mu \nu}
\label{} \eeq

\beq
B_{\mu \nu} = \frac{1}{1 + z_{\mu} z_{\nu}}(U^{-1})_{\mu x} (U^{-1})_{\nu x}
\label{} \eeq

\beq
\det( \hat{\Delta} - z \hat 1) = z^2 ( z -2 )^2 = 0
\label{} \eeq
The eigenvalues  $z_{\nu} $ are  following
\beqa
z_0& =& 0  \nonumber \\
z_1& =& 0  \nonumber \\
z_2& =& 2  \nonumber \\
z_3 &=& 2
\label{}
 \eeqa
As a result we have that for ${\bf q} = 0$ the function $B$ and
consequently $K$
has no singularity. Therefore we do not obtain the
phenomenon of "weak localization" in our 2D model
due to specific type of a disorder.

\section{Conclusions}
We have investigated the two-dimensional model with a new type
of disorder due to a random distribution of local
states with strongly anisotropic overlaps of wave functions.
The conductivity of this system was calculated in limits of
low and high densities of local states.
We have shown that considered type of disorder does not
lead to the weak localization phenomenon,
as usually in two-dimensional case \cite{rev},
due to the absence of logarithmic divergence from the integration
over the diffusion pole.

\section{Acknowledgments}
S.M. thanks A. Balatsky for a discussion.
 This work was supported by
the Russia  Fund  of Fundamental Research under grant No 960217791.

\end{document}